\documentclass{article}

\usepackage{microtype}
\usepackage{graphicx}
\usepackage{subfigure}
\usepackage{booktabs} 

\usepackage{enumitem} 
\usepackage{amsmath, amssymb, algorithmic, algorithm, amsfonts, array, color, bm, float, bbm, natbib}
\graphicspath{ {figures/} }
\DeclareMathOperator{\E}{E}
\DeclareMathOperator{\supp}{supp}
\DeclareMathOperator{\diag}{diag}

\newcolumntype{P}[1]{>{\centering\arraybackslash}p{#1}}
\newcolumntype{M}[1]{>{\centering\arraybackslash}m{#1}}

\usepackage{hyperref}

\usepackage[accepted]{noticml2018}

\begin{document}
	
\twocolumn[
\icmltitle{Black-box Variational Inference for Stochastic Differential Equations}



\icmlsetsymbol{equal}{*}

\begin{icmlauthorlist}
\icmlauthor{Thomas Ryder}{equal,stats,cs}
\icmlauthor{Andrew Golightly}{stats}
\icmlauthor{A.~Stephen McGough}{cs}
\icmlauthor{Dennis Prangle}{equal,stats}
\end{icmlauthorlist}

\icmlaffiliation{stats}{School of Mathematics, Statistics and Physics, Newcastle University, Newcastle, UK}
\icmlaffiliation{cs}{School of Computing, Newcastle University, Newcastle, UK}

\icmlcorrespondingauthor{Tom Ryder}{t.ryder2@newcastle.ac.uk}
\icmlcorrespondingauthor{Dennis Prangle}{dennis.prangle@newcastle.ac.uk}

\icmlkeywords{Machine Learning, ICML, SDEs, RNNs, Bayes, variational inference}

\vskip 0.3in
]



\printAffiliationsAndNotice{\icmlEqualContribution} 

\begin{abstract}
Parameter inference for stochastic differential equations is challenging due to the presence of a latent diffusion process.
Working with an Euler-Maruyama discretisation for the diffusion, we use variational inference to jointly learn the parameters and the diffusion paths.
We use a standard mean-field variational approximation of the parameter posterior, and introduce a recurrent neural network to approximate the posterior for the diffusion paths conditional on the parameters.
This neural network learns how to provide Gaussian state transitions which bridge between observations as the conditioned diffusion process does.
The resulting black-box inference method can be applied to any SDE system with light tuning requirements.
We illustrate the method on a Lotka-Volterra system and an epidemic model, producing accurate parameter estimates in a few hours.
\end{abstract}

\section{Introduction}

A stochastic differential equation (SDE) defines a \emph{diffusion process}, which evolves randomly over time, by describing its instantaneous behaviour.
As such, SDEs are powerful modelling tools used extensively in fields such as
econometrics \cite{Black:1973, Eraker:2001},
biology \cite{Gillespie:2000, Golightly:2011},
physics \cite{vanKampen:2007}
and epidemiology \cite{Fuchs:2013}.

It is only possible to work with analytic solutions to SDEs in special cases. Therefore it is common to use a numerical approximation, such as the Euler-Maruyama scheme.
Here the diffusion process is defined only on a grid of time points, and the transition density between successive diffusion states is approximated as Gaussian.
The approximation error involved converges to zero as the grid becomes finer.

Even under discretisation, statistical inference for SDEs observed at discrete times is challenging.
The difficulty is that, along with unknown parameters $\theta$ in the description of the SDE,
there is an unknown \emph{latent path} of the diffusion process, $x$.
An inference method must somehow deal with these high dimensional, highly structured latent variables.

Our proposed method uses recent advances in variational inference to jointly infer $\theta$ and $x$.
We introduce a flexible family of approximations to the posterior distribution and select the member closest to the true posterior.
We use a standard mean-field approximation for the $\theta$ posterior, and introduce a novel recurrent neural network (RNN) approximation for the posterior of $x$ conditional on $\theta$.
The RNN learns how to supply Gaussian state transitions between successive time points which closely match those for the intractable conditioned diffusion process.

Our black-box variational inference method is a simple and fast way to produce approximate inference for any SDE system.
We illustrate our method on Lotka-Volterra and epidemic examples, achieving accurate parameter estimates in just a few hours under default tuning choices.
Although our parameter posteriors are over-concentrated, as in most variational methods, our approximation of the conditioned diffusion process is close to the true posterior.
In comparison, existing Markov chain Monte Carlo (MCMC) methods (see Section \ref{sec:litreview}) require more tuning choices and can take days to run \cite{Whitaker:2017BA}.

\subsection{Related work} \label{sec:litreview}

\paragraph{Variational inference}

Several authors have looked at variational inference for SDEs \cite{Archambeau:2008} or related problems such as Markov jump processes \cite{Ruttor:2010} and state space models \cite{Archer:2016, Quiroz:2018}.
The novelty of our approach is to use:
(1) stochastic optimisation rather than variational calculus;
(2) a RNN-based variational approximation for the latent states instead of a mean-field or multivariate Gaussian approach.
We expect (2) is especially relevant to sparsely observed SDEs, where the latent states between observations may have a particularly complex dependency structure.


Another approach \cite{Moreno:2016} is to perform variational inference for the parameters only, using latent variables drawn from their prior in the ELBO estimate.
Such latent states are typically a poor match to the observed data and so make a negligible contribution to the ELBO.
To deal with the problem, close matches are upweighted.
Our approach avoids this extra approximation by instead learning the posterior distribution of the latent variables.

Our method can also be related to recent work on \emph{normalising flows} as variational approximations \cite{Rezende:2015}.
As in that work, our variational approximation can be viewed as transforming a $N(0,I)$ sample vector by successive linear transformations to an approximate posterior sample (of the diffusion states in our case).
Our work uses SDE theory to select simple and cheap transformations which produce a particularly good approximation.



\paragraph{Monte Carlo}

A popular approach in the Monte Carlo literature on SDEs is to introduce a \emph{bridge construct}:
an approximation to the discretised diffusion process conditional on the parameters and observations at a single time, derived using probability theory and various simplifying approximations.
The goal is to produce a path bridging between two observation times.
Combining successive bridges forms a complete diffusion path.
Bridge constructs can be used to produce proposals within Monte Carlo algorithms such as MCMC (see e.g.~\citealt{Roberts:2001, Golightly:2008, Fuchs:2013, vanderMeulen:2017}).
However, designing a bridge construct with desirable features for a particular problem is a challenging and time consuming tuning choice.
(Some particularly difficult regimes for bridge constructs are discussed in Section \ref{sec:experiments}.)
From this point of view, our contribution is to use machine learning to effectively automate the design of a bridge construct.

Another Monte Carlo approach is to perform approximate inference based on low dimensional summary statistics of the observations \cite{Picchini:2014}.
This results in a loss of information, which our approach avoids.

\section{Stochastic differential equations} \label{sec:SDEs}

Consider an It\^o process $\{X_t,t\geq0\}$ satisfying the SDE
\begin{equation} \label{eq:SDEs}
dX_t = \alpha(X_t, \theta) dt + \sqrt{\beta(X_t, \theta)} dW_t, \quad X_0 = x_0.
\end{equation}
Here $X_t$ is a $p$-dimensional vector of random variables, 
$\alpha$ is a $p$-dimensional \emph{drift vector}, $\beta$ is a $p \times p$ positive definite \emph{diffusion matrix} (with $\sqrt{\beta}$ representing a matrix square root) and $W_t$ is a $p$-vector of standard and uncorrelated Brownian motion processes.
The drift and diffusion depend on $\theta = (\theta_{1}, \theta_{2}, ... , \theta_{c})'$, a vector of unknown parameters (which may also include the initial condition $x_0$).

We assume that $\alpha(\cdot)$ and $\beta(\cdot)$ are sufficiently regular that \eqref{eq:SDEs} has a weak non-explosive solution \cite{øksendal2003stochastic}.
In this case, \eqref{eq:SDEs} defines a \emph{diffusion process}.
Such processes are always Markovian (i.e.~memoryless).

We further assume partial noisy observations of the latent process.
Suppose that there are $d+1$ observation times $t_0, t_1, \ldots, t_{d} = T$.
In the simplest case, these times are \emph{equally spaced}, separated by a time-step of $\Delta t$.
Let $y_{t_j}$ be a vector of $p_0$ observations at time $t_j$, for some $p_0 \leq p$.
Following \citet{Golightly:2008}, among others, we assume that
\begin{equation} \label{eq:observation model}
y_{t_j} = F' X_{t_j} + \omega_{t_j}, \quad \omega_{t_j} \overset{indep}{\sim } N(0, \Sigma),
\end{equation}
where $F$ is a constant $p \times p_0$ matrix,
and $\Sigma$ is a $p_0 \times p_0$ matrix which may be assumed known or the object of inference.
For the latter case $\Sigma$ should be a specified function of $\theta$.

Upon choosing a prior density $p(\theta)$, Bayesian inference proceeds via
the parameter posterior $p(\theta | y)$, or alternatively the joint posterior $p(\theta, x | y)$.

\paragraph{Discretisation}

Few SDEs permit analytical solutions to \eqref{eq:SDEs} and instead it is common to use an approximation based on time discretisation.
We therefore introduce intermediate time-points between observation times.
For concreteness, we present our methods for the case of equally spaced observations with $t_0=0$.
(It is easy to adapt them to alternative specifications of time points,
such as those required by irregularly-spaced observation times.)
We introduce $k-1$ time-points between successive observations,
giving a regular grid of times $\tau_i = i \Delta \tau$ for $i=0,1,2,\ldots,m=dk$,
with time-step $\Delta\tau=\Delta t/k$.
Note that $i=0,k,2k,\ldots, dk$ give the observation times.
The role of $k$ is to ensure the discretisation can be made arbitrarily accurate,
at the expense of increased computational cost.

We work with the simplest discretisation, the Euler-Maruyama scheme, in which transition densities between states at successive times are approximated as Gaussian
\begin{equation} \label{eq:EMT}
\begin{split}
p \left(x_{\tau_{i + 1}} | x_{\tau_{i}}, \theta\right) = \varphi \big(&x_{\tau_{i + 1}}-x_{\tau_{i}}; \, \alpha(x_{\tau_i}, \theta) \Delta \tau , \\
&\beta(x_{\tau_i}, \theta) \Delta \tau \big),
\end{split}
\end{equation}
where $\varphi (\cdot; \mu, S)$ is the Gaussian density with mean $\mu$ and variance matrix $S$.
A generative expression of this is
\begin{equation} \label{eq:EulerMaruyama}
x_{\tau_{i+1}} = x_{\tau_i} + \alpha(x_{\tau_i}, \theta) \Delta \tau + \sqrt{\beta(x_{\tau_i}, \theta) \Delta \tau}\, z_{i+1},
\end{equation}
where $z_{i+1}$ is an independent $N(0,I_p)$ realisation.

Discretisation is not guaranteed to preserve properties of the underlying SDE.
An issue which is particularly relevant later is positivity.
In many SDEs, such as population models, it is guaranteed that some components of $X_t$ are always positive.
However, in \eqref{eq:EulerMaruyama} $x_{\tau_{i+1}}$ is sampled from a Gaussian, which has unbounded support.
Consequently, there is a non-zero probability of sampling negative values.
This is problematic as the drift or diffusion function may be poorly behaved or undefined for such input.
A simple solution to this problem is the use of a reflecting boundary \cite{Skorokhod:1961}, for example by projecting invalid $x_{\tau_{i+1}}$ values back to the valid region \cite{Dangerfield:2012}.

\paragraph{Posterior}

The joint posterior under the Euler-Maruyama discretisation is
\begin{align}
p(\theta,x|y) &\propto p(\theta) p(x|\theta) p(y|x, \theta), \label{eq:posterior} \\
\begin{split}
p(x|\theta) &= 
\prod_{i=0}^{m-1} \varphi \big(x_{\tau_{i+1}} - x_{\tau_{i}};\, \alpha(x_{\tau_{i}},\theta) \Delta \tau,\,\\
&\qquad \quad \beta(x_{\tau_{i}}, \theta) \Delta \tau \big)
\end{split} \\
p(y|x, \theta) &= \prod_{i=0}^{d}\varphi\left(y_{t_i}; F' x_{t_i}, \Sigma \right).
\end{align}
In principle Monte Carlo algorithms can sample from \eqref{eq:posterior}.
However this is difficult in practice due to its high dimension and complex dependency structure.

\paragraph{Conditioned processes}

Consider the process defined by conditioning \eqref{eq:SDEs} on an initial state, $x_0$ and an exactly observed future state, $x_{t_1}$.
This conditioned process itself satisfies an SDE (see e.g.~\citealp{rogers_williams_2013}) with drift and diffusion
\begin{align}
\hat{\alpha}(x_t, \theta) &= \alpha(x_t, \theta)  + \beta(x_t, \theta) \nabla_{x_t}\log \pi\left(x_{t_1} | x_{t}, \theta\right), \label{eq:cond-sde} \\
\hat{\beta}(x_t, \theta) &= \beta(x_t, \theta), \label{eq:cond-sde2}
\end{align}
where $\pi\left(x_{t_1} | x_{t}, \theta\right)$ is the transition density of the unconditioned process.
While this is intractable in most cases, the result motivates our choice of variational approximation later.

In some simple situations a discretised approximation of this conditioned process can be derived (see e.g.~\citealt{Papaspiliopoulos:2013}) in which the diffusion matrix is scaled down as the observation time is approached.
Intuitively this is appealing: conditioned paths converge towards the observation, so nearby random deviations are smaller in scale.
This motivates us to use a variational approximation in which the diffusion matrix is not constrained to follow \eqref{eq:cond-sde2}, and instead is allowed to shrink.

\section{Approximate Bayesian inference}

Suppose we have a likelihood $p(y|\theta)$ for parameters $\theta$ under observations $y$.
Given a prior density $p(\theta)$ we wish to infer the posterior density $p(\theta|y) = p(\theta) p(y|\theta) / p(y)$.
It is typically possible to numerically evaluate the unnormalised posterior $p(\theta, y) = p(\theta) p(y|\theta)$.
Estimating the normalising constant $p(y) = \int p(\theta, y) d\theta$, known as the \emph{evidence}, is useful for Bayesian model selection.

\subsection{Variational inference}

Variational inference (VI) (see e.g.~\citealp{Blei:2017}) introduces a family of approximations to the posterior indexed by $\phi$, $q(\theta; \phi)$.
Optimisation is then used to find $\phi$ minimising the Kullback-Leibler divergence $KL(q(\theta; \phi) || p(\theta | y))$.
This is equivalent to maximising the ELBO (evidence lower bound) \cite{Jordan1999},
\begin{equation} \label{eq:ELBO}
\E_{\theta \sim q(\cdot; \phi)} [ \log p(\theta, y) - \log q(\theta;\phi) ].
\end{equation}
The optimal $q(\theta; \phi)$ is an approximation to the posterior distribution.
This is typically overconcentrated, unless the approximating family is rich enough to allow particularly close matches to the posterior.

The optimisation required by VI can be performed efficiently using the \emph{reparameterisation trick} \cite{journals/corr/KingmaW13, pmlr-v32-rezende14, icml2014c2_titsias14}.
This requires expressing $\theta \sim q(\cdot; \phi)$ as a \emph{non-centred parameterisation} \cite{Papaspiliopoulos:2003}.
That is, writing $\theta$ as the output of an invertible deterministic function $g(\epsilon, \phi)$ for some random variable $\epsilon$ with a fixed distribution.
Then the ELBO can be written as
\begin{equation} \label{eq:ELBO2}
\mathcal{L}(\phi) = \E_\epsilon [ \log p(\theta, y) - \log q(\theta;\phi) ],
\end{equation}
with an unbiased Monte-Carlo estimate 
\begin{equation} \label{eq:MonteCarloELBO}
\hat{\mathcal{L}}(\phi) = \frac{1}{n} \sum_{i=1}^n [ \log p(\theta^{(i)}, y) - \log q(\theta^{(i)};\phi) ],
\end{equation}
where $\theta^{(i)} = g(\epsilon^{(i)}, \phi)$ and
$\epsilon^{(1)}, \ldots, \epsilon^{(n)}$ are independent $\epsilon$ samples.
Assuming $\hat{\mathcal{L}}$ is differentiable with respect to $\phi$,
the gradient of \eqref{eq:MonteCarloELBO} can be calculated using automatic differentiation,
and the resulting unbiased estimator of $\nabla \mathcal{L}(\phi)$ used in stochastic gradient descent or similar algorithms.

\subsection{Importance sampling}

When variational inference outputs a good match to the posterior distribution,
\emph{importance sampling} (IS) (see e.g.~\citealp{Robert:2004}) can correct remaining inaccuracies and provide near-exact posterior inference.
In more detail, select an \emph{importance density} $q(\theta)$ which can easily be sampled from,
and satisfies $\supp q(\theta) \supseteq \supp p(\theta | y)$.
IS samples $\theta^{(1)}, \theta^{(2)}, \ldots, \theta^{(N)}$ from $q$ and calculates weights $w_i = p(\theta^{(i)},y)/q(\theta^{(i)})$.
Then, for any function $h$, an estimate of $\E_{\theta \sim p(\cdot|y)}[h(\theta)]$ is
\begin{equation} \label{eq:IS}
\sum_{i=1}^N h(\theta^{(i)}) w_i \bigg/ \sum_{i=1}^N w_i.
\end{equation}
This is consistent for large $N$, but in practice $q$ should approximate the posterior for accurate estimation at a feasible cost.
Also note that $N^{-1} \sum_{i=1}^N w_i$ is an unbiased and consistent estimate of the evidence.

A diagnostic for the quality of IS results is the \emph{effective sample size} (ESS).
This is defined as
\begin{equation} \label{eq:ESS}
N_{\text{eff}} = \left(\sum_{i=1}^N w_i\right)^2 \bigg/ \ \sum_{i=1}^N w_i^2.
\end{equation}
For most functions $h$, the variance of \eqref{eq:IS} approximately equals that of an idealised Monte Carlo estimate based on $N_{\text{eff}}$ independent samples from $p(\theta|y)$ \cite{Liu:1996}.
In practice we will use a variational approximation as the importance density,
and the ESS to assess whether this is sufficiently good to produce accurate estimates.
However, ESS values can be an unstable for poor importance densities \cite{Vehtari:2017} so later we also consider other problem-specific evidence for the quality of our results.


\section{Variational inference for SDEs}

Our variational approximation to the posterior \eqref{eq:posterior} is
\begin{equation} \label{eq:variationalApprox}
q(\theta, x; \phi) = q(\theta; \phi_\theta) q(x | \theta; \phi_x).
\end{equation}
These factors represent approximations to $p(\theta | y)$ and $p(x | \theta, y)$ respectively,
which are described below.
Here $\phi_\theta$ and $\phi_x$ are the variational parameters for the two factors,
and $\phi$ is the collection of all variational parameters.
Note our eventual choice of $q(x | \theta; \phi_x)$ depends on several features of the data and model (see list in Section \ref{sec:impldetails}), but we suppress this in our notation for simplicity.

\subsection{Approximate parameter posterior}

For $q(\theta; \phi_\theta)$ we use the \emph{mean-field Gaussian} approximation
\begin{equation} \label{eq:MFG}
q(\theta; \phi_\theta) = \prod_{i=1}^c \varphi(\theta_i; \mu_i, s_i^2),
\end{equation}
with $\phi_\theta=(\mu_1, \ldots, \mu_c, s_1, \ldots, s_c)$.
Hence the components of $\theta$ are independent Gaussians.
To express $\theta$ using a non-centred parameterisation, we write
\begin{equation} \label{eq:theta}
\theta = g_\theta(\epsilon_1, \phi_\theta) = S \epsilon_1  + \mu.
\end{equation}
where $\epsilon \sim N(0,I_c)$, $S = \diag(s_1, \ldots, s_c)$ and $\mu = (\mu_1, \ldots, \mu_c)$.

It may be necessary to transform $\theta$ to an alternative parameterisation $\vartheta$ so that a Gaussian approximation is appropriate e.g.~log-transforming parameters constrained to be positive.

Mean-field approximations are imperfect, often producing underdispersed estimates of the posterior (see e.g.~\citealp{Blei:2017}),
and more sophisticated approximations (e.g.~\citealp{Rezende:2015}) could be used here instead.
However mean-field approximations suffice to give good parameter estimation in our examples.

\subsection{Approximate conditioned diffusion process}
Motivated by the result that a diffusion process conditioned on an exact observation is itself an SDE (see Section \ref{sec:SDEs}),
we base $q(x|\theta; \phi_x)$ upon a discretised diffusion.
A generative definition is
\begin{equation} \label{eq:VariationalDist}
\begin{split}
x_{\tau_{i+1}} = h \bigg(& x_{\tau_i} + \tilde{\alpha}(x_{\tau_i}, y, \theta, \tau_i; \phi_x) \Delta\tau \\
&+ \sqrt{\tilde{\beta}(x_{\tau_i}, y, \theta, \tau_i; \phi_x) \Delta\tau} z_{i+1} \bigg),
\end{split}
\end{equation}
where $\tilde{\alpha}$ and $\tilde{\beta}$ are drift and diffusion functions.
Taking $h$ as the identity function gives a discretised diffusion process.
However often we use $h$ to impose positivity constraints on some components of $x$ -- see Section \ref{sec:softplus}.

The resulting variational density $q(x| \theta; \phi_{x})$ is
\begin{equation}
\begin{split}
\prod_{i=0}^{m-1} \varphi\bigg(&x_{\tau_{i+1}} - x_{\tau_{i}};
\tilde{\alpha}(x_{\tau_{i}}, y, \theta, \tau_i; \phi_{x}) \Delta\tau,\, \\
& \tilde{\beta}(x_{\tau_{i}}, y, \theta, \tau_i; \phi_{x}) \Delta\tau\bigg)
|\det J_i|.
\end{split}
\end{equation}
where $J_i$ is the Jacobian matrix associated with the transformation $h$ in \eqref{eq:VariationalDist}.
To express $x$ with a non-centred parameterisation, let $\epsilon_2 \sim N(0,I_{pm})$ be the flattened vector of $(z_1, z_2, \ldots, z_m)$ realisations.
Then apply \eqref{eq:VariationalDist} repeatedly.
Let the outcome be represented by the function
\begin{equation} \label{eq:x}
x = g_x(\epsilon_2, \theta, \phi_x).
\end{equation}

We use a neural network, with parameters $\phi_x$, to serve as our functions $\tilde{\alpha}$ and $\tilde{\beta}$.
At time $\tau_i$ it acts as follows.
The network's input is several features
(described in Section \ref{sec:impldetails})
computed from: the current diffusion state $x_{\tau_i}$, the observations $y$, the parameters $\theta$ and the current time $\tau_i$.
The network outputs a drift vector and diffusion matrix
(see Section \ref{sec:impldetails} for details of the latter), which are used to sample $x_{\tau_{i+1}}$ from \eqref{eq:VariationalDist}.
This state forms part of the neural network input at time $\tau_{i+1}$.
So the network just discussed forms a \emph{cell} of an overall recurrent neural network (RNN) structure for $q(x| \theta; \phi_x)$.
Note that long-term memory features are not required as we wish to produce a diffusion process, which is memoryless.

\subsection{Ensuring positivity} \label{sec:softplus}

In practice, we take $h$ in \eqref{eq:VariationalDist} to be
a function which applies the identity function to unconstrained components of the diffusion state,
and the function $\text{softplus}(z) = \log(1+e^z)$ to components with positivity constraints.
This function produces strictly positive outputs while having little effect on positive inputs above 2.
The latter property means our variational approximation usually remains similar to a discretised diffusion process.
Alternative transformations could be used if state values below 2 were believed to be common, potentially with tuning parameters so a suitable shape could be learned.
However, this was not necessary for our examples.

Preliminary work found this transformation approach to enforcing positivity was much easier to implement in the variational framework than reflection methods.
It can be interpreted as constraining the variational approximation based on prior beliefs about positivity of diffusion paths.

\subsection{Algorithm}

Our overall inference algorithm is given in Algorithm \ref{alg:VIforSDEs}.
This aims to maximise the ELBO
\begin{equation} \label{eq:ELBO3}
\mathcal{L}(\phi) = \E_{\theta, x \sim  q(\cdot; \phi)}\left[ \log \frac{ p (\theta) p(x|\theta) p (y|x, \theta)}{ q(\theta; \phi_{\theta}) q(x | \theta ; \phi_{x})}\right],
\end{equation}
by differentiating the Monte Carlo estimate
\begin{equation} \label{eq:sgvb}
\hat{\mathcal{L}}(\phi) = \frac{1}{n} \sum_{i=1}^n \log \frac{ p (\theta^{(i)}) p(x^{(i)}|\theta^{(i)}) p (y|x^{(i)}, \theta^{(i)})}{ q\left(\theta^{(i)}; \phi_{\theta}\right) q\left(x^{(i)} | \theta^{(i)}; \phi_x\right)},
\end{equation}
where $\theta^{(i)} = g_\theta(\epsilon_1^{(i)}, \phi_\theta)$
and $x^{(i)} = g_x(\epsilon_2^{(i)}, \theta^{(i)}, \phi_x)$.

\begin{algorithm}[htb]
\caption{Black-box variational inference for SDEs}
\label{alg:VIforSDEs}
\begin{algorithmic}
	\STATE Initialise $\phi_0$ and $k=0$.
	\LOOP
	\STATE Sample $\epsilon_1^{(i)}, \epsilon_2^{(i)}$ for $1 \leq i \leq n$.
	\STATE Calculate $\nabla \hat{\mathcal{L}}(\phi_k)$ using automatic differentiation of \eqref{eq:sgvb}.
	\STATE Calculate $\phi_{k+1}$ using stochastic gradient descent, or a similar algorithm, and increment $k$.
	\ENDLOOP
\end{algorithmic}
\end{algorithm}

\subsection{Implementation details} \label{sec:impldetails}

Our RNN cell input at time $\tau_i$, with $t_j \leq \tau_i < t_{j+1}$, is:
\begin{itemize}[noitemsep]
\item The parameters $\theta$.
\item The most recent latent state, $x_{\tau_{i-1}}$.
\item The time until the next observation, $t_{j+1} - \tau_i$.
\item The next observation time, $t_{j+1}$.
\item The difference between the next observation and what the mean observation would be at the most recent latent state, $y_{t_{j+1}} - F' x_{\tau_{i-1}}$.
\end{itemize}
Exploratory work showed that the RNN produces a much better approximation of the conditioned process
with these features as input rather than simply $x_{\tau_i}, y, \theta$ and $\tau_i$.

Our RNN cell outputs a vector $\tilde{\alpha}$ and the coefficients of a lower-triangular matrix, $M$. 
In order to return a Cholesky factor of $\tilde{\beta}$, the diagonal elements of $M$ are transformed using the softplus function to ensure positivity.
We also regularise to avoid $\tilde{\beta}$ matrices with very small determinants.

Algorithm \ref{alg:VIforSDEs} requires automatic differentiation of \eqref{eq:sgvb}.
This can be achieved using the standard tool-kit of backpropagation after \emph{rolling-out} the RNN
i.e.~stacking $m$ copies of the RNN cell to form a deep feed-forward network.
The canonical challenge in training such networks, known as the exploding-gradient problem (\citealp{Bengio:1994:LLD:2325857.2328340}), often necessitates the use of gradient clipping to control for numerical instability.
We follow \citet{Pascanu:2013:DTR:3042817.3043083} and perform gradient clipping using the $L_1$ norm.

To initialise $\phi_0$ in Algorithm \ref{alg:VIforSDEs}, we select $\phi_\theta$ so the margins of the variational approximation are based on those of the parameter priors.
Standard choices from the neural network literature -- random Gaussian weights and constant biases -- are used for $\phi_x$.


\section{Experiments} \label{sec:experiments}

We implement our method for two examples:
(1) analysing synthetic data from a Lotka-Volterra SDE; (2) analysing real data from an SDE model of a susceptible-infectious-removed (SIR) epidemic.
Our experiments include challenging regimes such as: (A) low-variance observations; (B) conditioned diffusions with non-linear dynamics; (C) unobserved time series; (D) widely spaced observation times; (E) data which is highly unlikely under the unconditioned model.
Many of these violate the assumptions used by existing diffusion bridge constructs \cite{Whitaker:2017SC}.

In all our experiments below similar tuning choices worked well.
We use batch size $n=50$ in \eqref{eq:sgvb}.
Our RNN cell has four hidden layers each with 20 hidden units and rectified-linear activation.
We implement our algorithms in Tensorflow using the Adam optimiser \cite{DBLP:journals/corr/KingmaB14} and report results using an 8-core CPU.
The code is available at \url{https://github.com/Tom-Ryder/VIforSDEs}.

\subsection{Lotka-Volterra} \label{sec:LV}

Lotka-Volterra models describe simple predator-prey population dynamics combining three types of event: prey reproduction, predation (in which prey are consumed and predators have the resources to reproduce) and predator death.
A SDE Lotka-Volterra model (for derivation see e.g.~\citealp{Golightly:2011}) is defined by
\begin{align}
\alpha(X_t, \theta) &=
\begin{pmatrix}
\theta_1 U_t  - \theta_2 U_tV_t\\
\theta_2 U_t V_t - \theta_3 V_t
\end{pmatrix}, \\ 
\beta(X_t, \theta) &=
\begin{pmatrix}
\theta_1 U_t  +\theta_2 U_tV_t  &-\theta_2 U_tV_t\\
-\theta_2 U_tV_t & \theta_3 V_t+\theta_2U_t V_t 
\end{pmatrix},
\end{align}
where $X_t = (U_t, V_t)'$ represents the populations of prey and predators at time $t$.
The parameters $\theta = \left(\theta_1, \theta_2, \theta_3\right)'$ control the rates of the three events described above.

\begin{figure}[htb] 
  \centering
  \includegraphics[width = .48\textwidth]{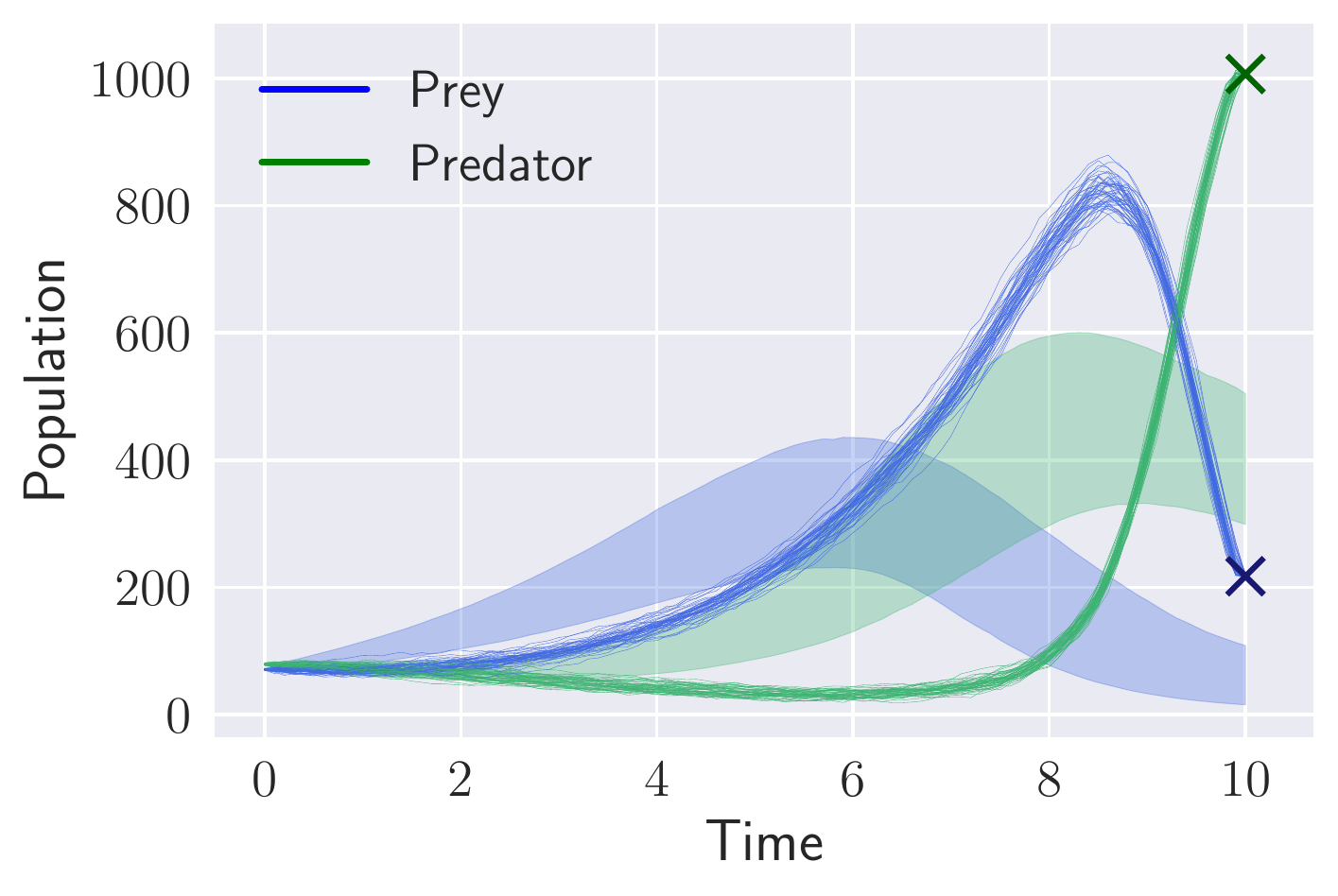}
  \caption{50 conditioned diffusion paths for the Lotka-Volterra example with a single observation time and known parameters.
    These are sampled from the trained variational approximation.
    The observations used are from the last row of Table \ref{tab:LV}, representing highly unlikely observations.
    The shaded regions show typical paths of the unconditioned process
(95\% intervals estimated from repeated simulations.)}
\label{fig:LV_single_bridge}
\end{figure}

In the experiments below, we use discretisation time step $\Delta \tau = 0.1$ and observation variance $\Sigma=I$, which is small relative to the typical population sizes (see e.g.~Figure \ref{fig:LV_single_bridge}.)

\paragraph{A single observation time with known parameters} \label{sec:LV_single_bridge} 
We begin with the case of a single observation time and known parameter values, where we follow \citet{Boys2008} by taking $\theta = \left(0.5, 0.0025, 0.3\right)'$ and $x_{0} = (71, 79)'$.
This setting solely investigates our ability to learn $x$, without uncertainty in $\theta$: essentially the same problem as creating a bridge construct (described in Section \ref{sec:litreview}.)

We consider four different observations at time $t=10$, listed in Table \ref{tab:LV}.
For each example we train our variational approximation until convergence (assessed manually throughout this paper),
which takes roughly 20 minutes for the first 3 examples, and 90 minutes for the last.
We then perform importance sampling using 500,000 samples from the fitted approximation.
Table \ref{tab:LV} shows the resulting ESS values.
The first 3 rows in the table are typical observations under the model given our $\theta$, while the final row represents highly unlikely observations (double those in the previous row).
Figure \ref{fig:LV_single_bridge} shows fitted diffusion paths for this case.

\begin{table}[htp]
\caption{Summary of importance sampling performance for the Lotka-Volterra example with a single observation time and known $\theta$.
Each row shows the prey ($U_{10}$) and predator ($V_{10}$) observations at time $t=10$.
For each, 500,000 iterations of importance sampling are performed,
using variational inference output as the importance density,
and the effective sample size \eqref{eq:ESS} is shown.
}\label{tab:LV}
\centering
\begin{tabular}{|cc|c|} 
\hline
$U_{10}$ & $V_{10}$ & ESS     \\
\hline
15.3     & 298.2    & 184,329 \\
46.7     & 389.1    & 212,313\\
108.7    & 503.4    & 196,956 \\
217.4    & 1006.9   & 95,711 \\
\hline
\end{tabular}
\end{table}

This example contains several challenging features: all those listed at the start of Section \ref{sec:experiments} except (C).
While these features make it hard to use existing bridge constructs, our variational method produces a close approximation to the true posterior, as illustrated by the high ESS values.

The case of highly unlikely observations takes longer to train and receives a lower ESS,
reflecting that a more complicated diffusion path must be learned here.
(To check this we found that a simpler RNN cell suffices for good performance in the other examples but not this one.)

\begin{figure}[htb]
	\centering
	\includegraphics[width = .48\textwidth]{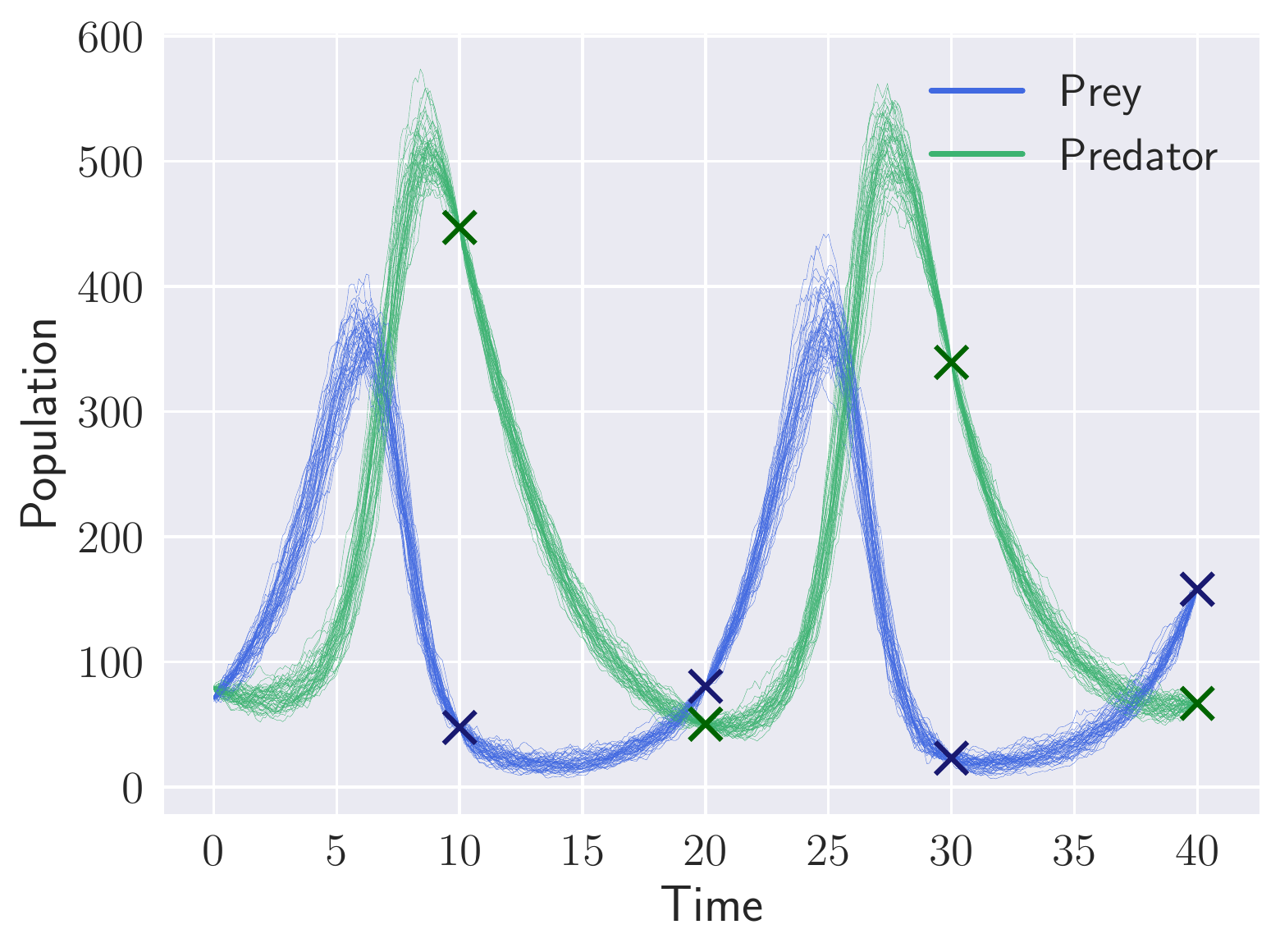}
	\caption{50 conditioned diffusion paths for the Lotka-Volterra example with multiple observation times and unknown parameters. They are sampled from the fitted variational approximation.}
	\label{fig:LV_sparse_bridges}
\end{figure}

\paragraph{Multiple observation times with known parameters}
We now extend the previous example to multiple observation times, $t=0,10,20,30,40$.
We analyse synthetic data, produced using the parameters specified previously (including observation noise with $\Sigma=I$).
Here convergence takes 6 hours, and importance sampling with 500,000 samples produces an ESS of 96,812.
The resulting diffusion paths are not shown as they are very similar visually to the next example (see Figure \ref{fig:LV_sparse_bridges}).

This example shows that our method learns the conditioned process well even when there are several observation times.

\paragraph{Multiple observation times with unknown parameters}
We now analyse the same synthetic data with unknown $\theta$ parameters.
As these must take positive values, we work with the log-transformed parameters $\vartheta$ and assume they have independent $N(0, 3^2)$ priors.
Our results are shown after transforming back to the original parameterisation.

\begin{figure}[htb]
	\centering
	\includegraphics[width = .48\textwidth]{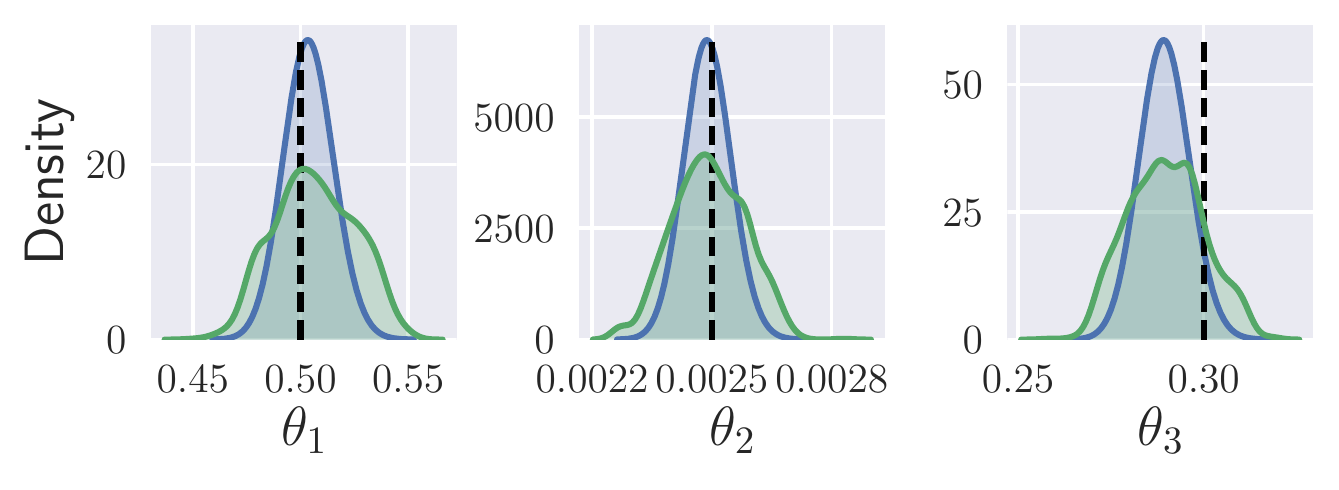}
	\caption{Marginal posterior density estimates for the Lotka-Volterra example from variational inference (blue) and importance sampling (green). Black dashed lines show true parameter values.}
	\label{fig:LV_params}
\end{figure}

Convergence takes 2 hours, and importance sampling with 500,000 iterations produces an ESS of 635.4.
Figure \ref{fig:LV_sparse_bridges} shows 50 diffusion paths sampled from the fitted variational approximation.
Figure \ref{fig:LV_params} shows two estimates of the marginal parameter posteriors:
the variational inference output, and a kernel density estimate based on importance sampling results.

The estimated posteriors give accurate estimates of the true parameter values.
However, the low ESS here shows that both estimates of the parameter posteriors are imperfect approximations (also illustrated by the variational posterior estimates appearing overconcentrated compared to the importance sampling results.)
Achieving good point estimates but imperfect posteriors is typical for variational inference \cite{Blei:2017}.

\subsection{Epidemic model} \label{sec:epi}
An SIR epidemic model \cite{Andersson:2000} describes the spread of an infectious disease.
The population is split into those susceptible ($S$), infectious ($I$) and removed ($R$).
Two types of event take place: susceptibles can be infected by the infectious, and the infectious eventually become removed.
Constant population size is assumed.
Hence only the $S$ and $I$ population sizes need to be modelled.

An SIR epidemic model using SDEs is defined by
\begin{align}\label{eq:SIR}
\alpha(X_t, \theta) & = \begin{pmatrix}
- \theta_1 S_{t} I_{t} \\
\theta_1 S_{t} I_{t} - \theta_2 I_{t},
\end{pmatrix} \\
\beta(X_t, \theta) &= \begin{pmatrix}
\theta_1 S_{t} I_{t} & -\theta_1 S_{t} I_{t}\\
- \theta_1 S_{t} I_{t} & \theta_1 S_{t} I_{t} + \theta_2 I_{t} 
\end{pmatrix},
\end{align}
where $X_t = (S_t, I_t)' $ is the state of the system at time $t$,
$\theta_1$ is an infection parameter and $\theta_2$ is a removal parameter.
For a detailed derivation see \citet{Fuchs:2013}.

Our data is taken from an outbreak of influenza at a boys boarding school in 1978 (\citealp{Jacksone002149}).
Influenza was introduced to the population by a student returning from holiday in Hong Kong. 
Of the 763 boys at the school, 512 were infected within 14 days.
Hence we assume $x_0 = (762, \, 1)'$.
Observations of the number infectious are provided daily by those students confined to bed.
We assume Gaussian observation error with unknown variance $\sigma^2$.
Our analyses use a discretisation time step of $\Delta \tau = 0.1$.

We also consider an alternative model with a time-varying infection parameter.
Here we let $\vartheta_1 = \log \theta_1$ follow an Ornstein-Uhlenbeck process
\begin{equation}\label{eq:OU}
d \vartheta_{t,1} = \theta_{3}\left(\theta_{4} - \vartheta_{t,1}\right)dt + \theta_{5} dW_{t},
\end{equation}
where $\theta_3$, $\theta_4$ and $\theta_5$ are the mean-reversion rate, process mean and volatility, respectively, and $\theta_{t,1}$ is the infection parameter at time $t$.
Previous related work has focused on ODE epidemic models with time-varying parameters following SDEs \cite{Dureau:2013, delMoral:2015}.
In contrast, our approach can easily be applied to a full SDE system.

\paragraph{Time-invariant infection parameter}
We infer the log-transformed parameters
$\vartheta = (\log \theta_1, \log \theta_2, \log \sigma^2)'$
under independent $N(0, 3^2)$ priors.
Our results are shown after transforming back to the original parameterisation.

\begin{figure}[htp]
	\centering
	\includegraphics[width = .48\textwidth]{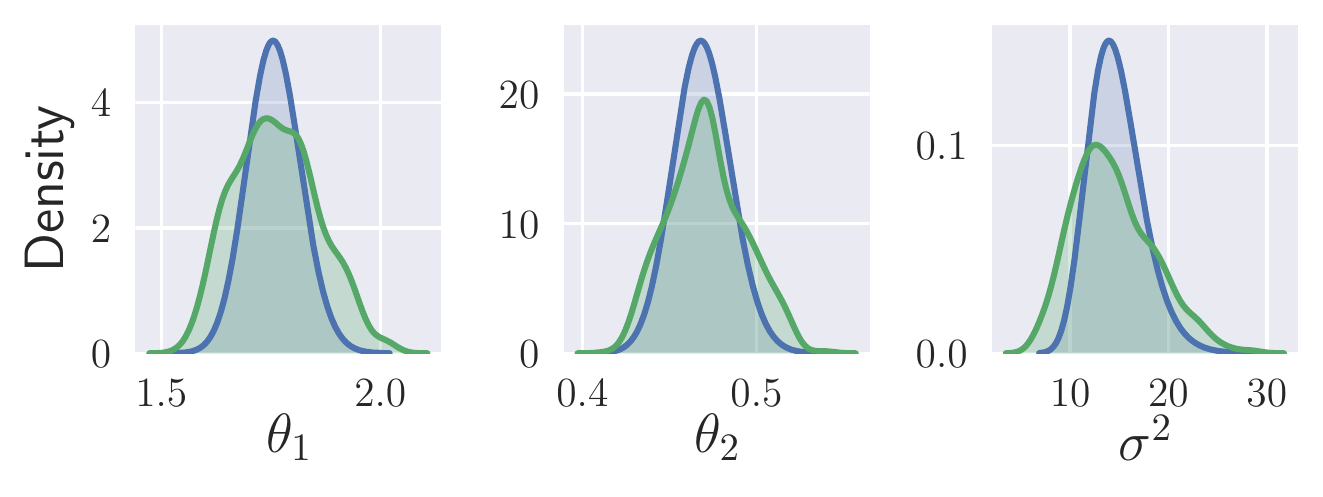}
	\caption{Marginal posterior density estimates from variational inference (blue) and importance sampling (green) for the SIR epidemic model with constant $\theta_1$.}
	\label{fig:SIR_invariant_params}
	\includegraphics[width = .48\textwidth]{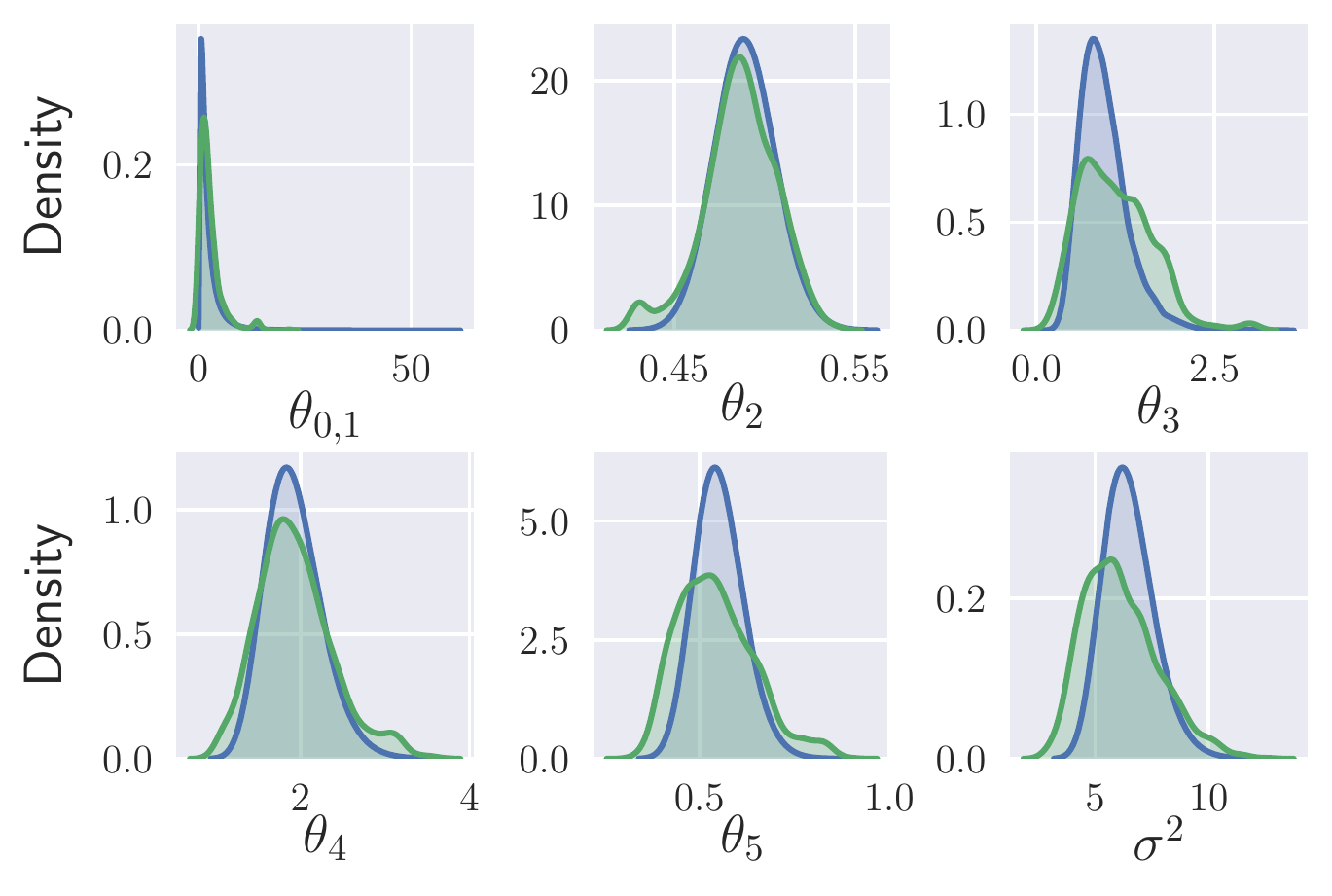}
	\caption{Marginal posterior density estimates from variational inference (blue) and importance sampling (green) for the SIR epidemic model with time varying $\theta_1$.}
	\label{fig:SIR_variant_params}
\end{figure}

\begin{figure}[htp] 
	\centering
	\includegraphics[width = 0.48\textwidth]{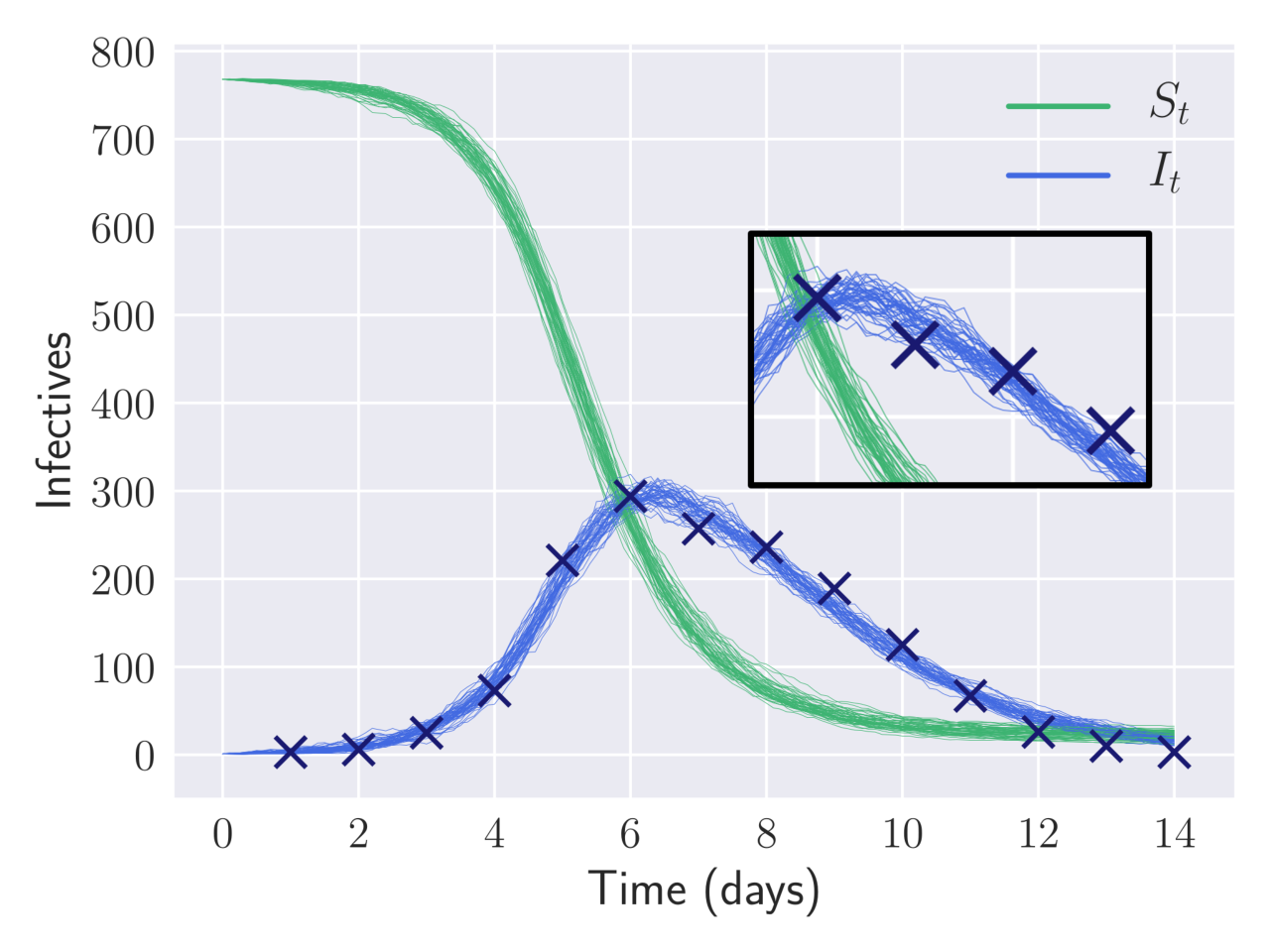}
	\caption{50 conditioned diffusion paths for the SIR epidemic model with constant $\theta_1$, sampled from the trained variational approximation.
          The observations are represented by dark blue crosses.}
	\label{fig:SIR_invariant_bridges}
	\includegraphics[width = 0.48\textwidth]{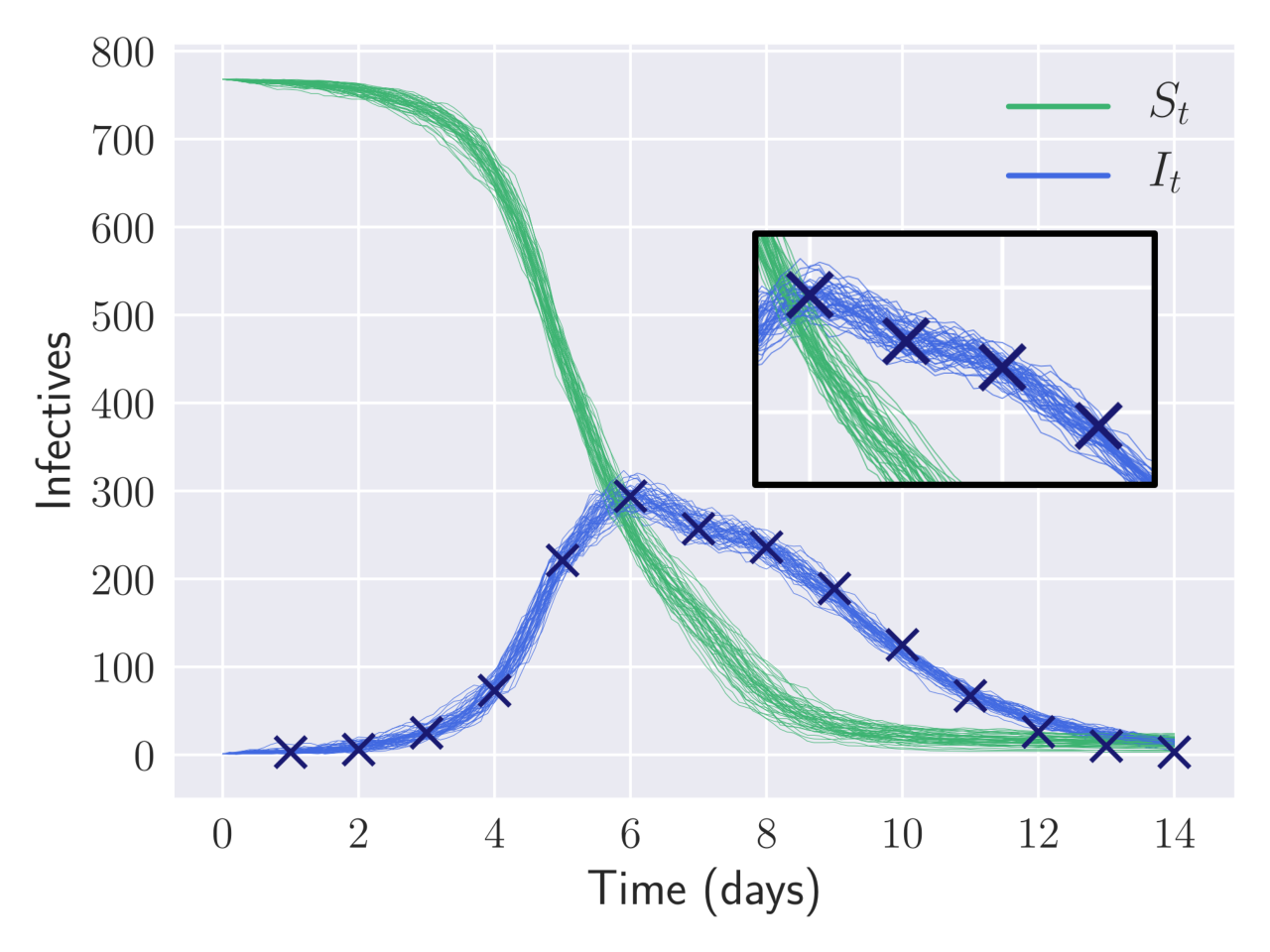}
	\caption{50 conditioned diffusion paths for the SIR epidemic model with time varying $\theta_1$, sampled from the trained variational approximation.
          The observations are represented by dark blue crosses.}
	\label{fig:SIR_variant_bridges}
	\includegraphics[width = .48\textwidth]{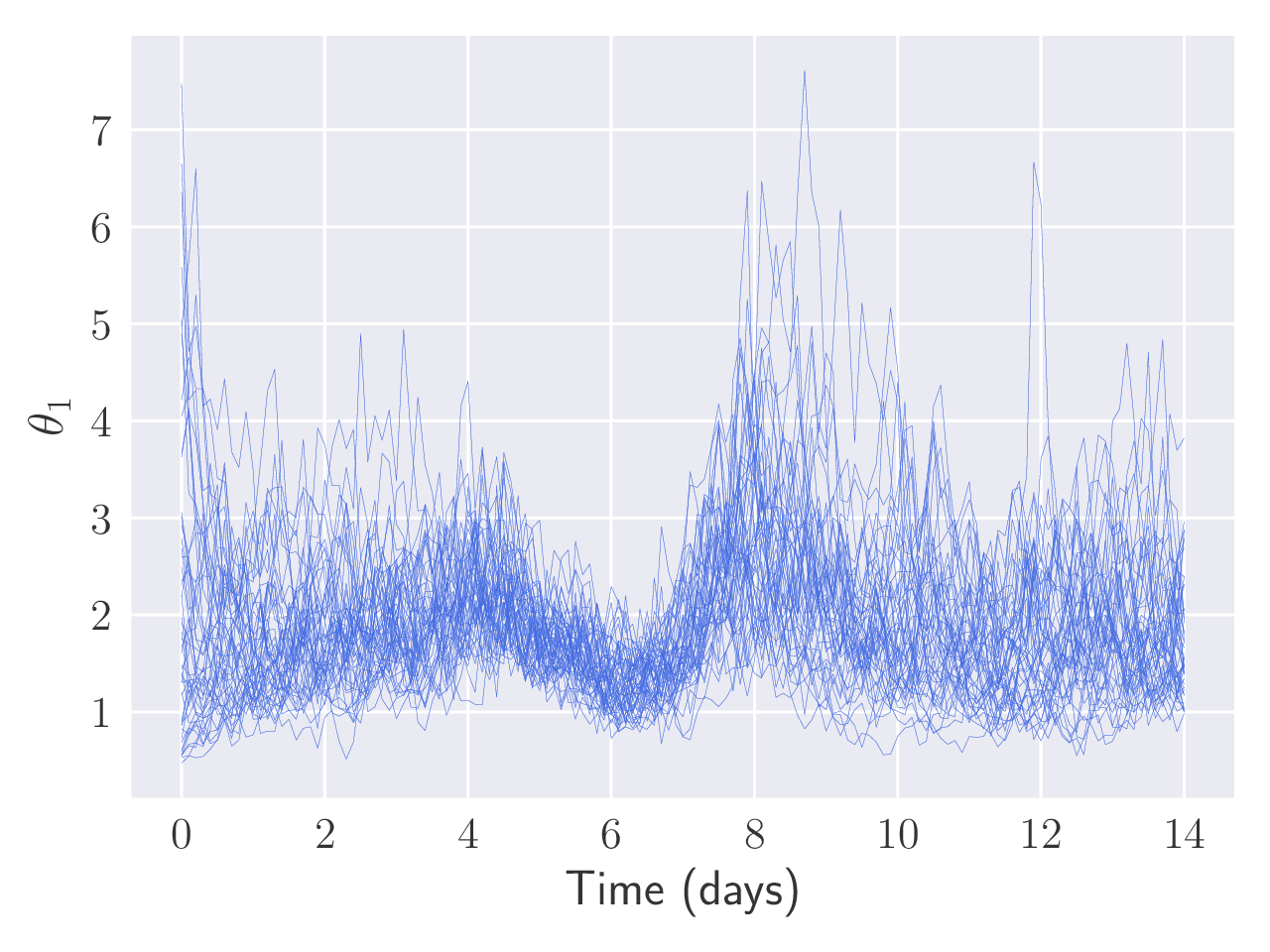}
	\caption{50 conditioned diffusion paths for $\theta_1$ for the SIR epidemic model, sampled from the trained variational approximation.}
	\label{fig:OU_bridges}
\end{figure}

Convergence takes 2.5 hours, and importance sampling with 500,000 iterations produces an ESS of 718.2.
Figure \ref{fig:SIR_invariant_params} shows two estimates of the marginal parameter posteriors:
variational inference output, and a kernel density estimate based on importance sampling results.
Figure \ref{fig:SIR_invariant_bridges} shows 50 diffusion paths sampled from the variational approximation.

The small ESS indicates there is some approximation error.
However, the marginal posteriors for $\theta_1$ and $\theta_2$ are very similar to those from the MCMC analysis of \citet[pg 293]{Fuchs:2013}, despite some modelling differences (that analysis fixed $\sigma^2=0$ and used exponential priors for $\theta_1$ and $\theta_2$).

\paragraph{Time-variant infection parameter}
We infer the log-transformed parameters
$\vartheta = (\log \theta_{0,1},\log \theta_2,\log \theta_3,\log \theta_4,$ $\log \theta_5, \log \sigma^2)'$
under independent $N(0, 3^2)$ priors.
Results are shown after transforming back to the original parameterisation.

Convergence now takes 3 hours, and 500,000 iterations of importance sampling produces an ESS of 256.1.
Figure \ref{fig:SIR_variant_params} shows estimates of the marginal parameter posteriors, using variational inference and importance sampling outputs as before.
Figures \ref{fig:SIR_variant_bridges} (SIR) and \ref{fig:OU_bridges} (Ornstein-Uhlenbeck) show 50 diffusion paths sampled from the variational approximation.
Again the low ESS indicates some approximation error.

\paragraph{Model comparison}
The two models produce visually similar diffusion paths, but close inspection shows some differences.
The time-invariant model paths for $I_t$ appear smooth but slightly miss some of the observation points.
The time-variant model paths for $I_t$ are less smooth and more accurately capture the shape of the data.
Correspondingly, the time-varying model infers a smaller $\sigma^2$ value.
The most obvious difference in $I_t$ paths occurs for the $t=7,8,9$ observations, shown in the zoomed-in inset of Figures \ref{fig:SIR_invariant_bridges} and \ref{fig:SIR_variant_bridges}.
Figure \ref{fig:OU_bridges} shows that shortly before this the time-variant $\theta_1$ values become constrained to smaller values.

Although the time-varying model appears to fit the data better, this is at the cost of increased model complexity, and could simply reflect overfitting.
A better estimate of the parameter posteriors would allow formal model comparison based on importance sampling evidence estimates.


\section{Conclusion}

We provide a black-box variational approach to inference for SDES which is simple, practical and fast (relative to existing methods).
This performs inference for a broad class of SDEs with minimal tuning requirements.
Empirical investigation shows we obtain close matches to the posterior of the conditioned diffusion paths.
Approximate parameter inference is also possible, with our results recovering known parameters for synthetic data (Section \ref{sec:LV}), and previous results for real data (Section \ref{sec:epi}), using only a few hours of computation for a desktop PC.
An interesting future direction is develop choices of $q(x|\theta;\phi_x)$ more efficent than standard RNNs, to further reduce computing time and enable real-time applications of this methodology.

\newpage
\section*{Acknowledgements}

Tom Ryder is supported by the Engineering and Physical Sciences Research Council, Centre for Doctoral Training in Cloud Computing for Big Data (grant number EP/L015358/1).

We acknowledge with thanks an NVIDIA academic GPU grant for this project.

\bibliography{VIforSDEs}
\bibliographystyle{icml2018}

\end{document}